\documentclass[conference]{IEEEtran}
\IEEEoverridecommandlockouts

\usepackage[ruled,vlined,lined,commentsnumbered]{algorithm2e}
\usepackage{booktabs}
\usepackage{moreverb}
\usepackage{fontenc}
\usepackage{amsmath,amsfonts}
\usepackage{amssymb}
\usepackage{fancybox}
\usepackage{color}
\usepackage{colortbl}
\usepackage{array}
\usepackage{multirow}
\usepackage{multicol}
\usepackage{listings}
\usepackage{makecell}
\usepackage{graphicx}

\usepackage[ruled,vlined,lined,commentsnumbered]{algorithm2e}
\usepackage{balance}
\usepackage{csvsimple}
\usepackage{longtable}
\usepackage{url}
\usepackage{lscape}
\usepackage{rotating}
\usepackage{tikz}
\usepackage[normalem]{ulem}
\usepackage{enumitem}
\usepackage{cite}

\usepackage[most]{tcolorbox}

\usepackage{threeparttable}

\usepackage{marvosym}
\usepackage{pifont}

\newcolumntype{L}[1]{>{\raggedright\let\newline\\\arraybackslash\hspace{0pt}}m{#1}}
\newcolumntype{C}[1]{>{\centering\let\newline\\\arraybackslash\hspace{0pt}}m{#1}}
\newcolumntype{R}[1]{>{\raggedleft\let\newline\\\arraybackslash\hspace{0pt}}m{#1}}

\definecolor{codegreen}{rgb}{0,0.6,0}
\definecolor{codegray}{rgb}{0.5,0.5,0.5}
\definecolor{codepurple}{rgb}{0.58,0,0.82}
\definecolor{backcolour}{rgb}{0.95,0.95,0.92}

\lstdefinestyle{mystyle}{
    commentstyle=\color{codegreen},
    keywordstyle=\color{magenta},
    numberstyle=\tiny\color{codegray},
    stringstyle=\color{codepurple},
    basicstyle=\footnotesize,
    breakatwhitespace=false,
    breaklines=true,
    captionpos=b,
    keepspaces=true,
    showspaces=false,
    showstringspaces=false,
    showtabs=false,
    tabsize=2
}

\setlist{noitemsep} 

\definecolor{darkpastelred}{rgb}{0.76, 0.23, 0.13}
\definecolor{ao(english)}{rgb}{0.0, 0.5, 0.0}

\definecolor{darkpastelred}{rgb}{0.76, 0.23, 0.13}
\definecolor{ao(english)}{rgb}{0.0, 0.5, 0.0}

\lstdefinelanguage{diff}{
  morecomment=[f][\color{blue}]{@@},     
  morecomment=[f][\color{red}]-,         
  morecomment=[f][\color{codegreen}]+,       
  morecomment=[f][\color{red}]{---}, 
  morecomment=[f][\color{codegreen}]{+++},
}

\definecolor{yellow}{RGB}{255,255,153}
\definecolor{grey}{RGB}{224,224,224}

\newboolean{showcomments}
\setboolean{showcomments}{true}
\ifthenelse{\boolean{showcomments}}
 { \newcommand{\mynote}[2]{
      \fbox{\bfseries\sffamily\scriptsize#1}
        {\small$\blacktriangleright$\textsf{\emph{#2}}$\blacktriangleleft$}}}
        { \newcommand{\mynote}[2]{}}

\definecolor{DarkOrange}{rgb}{0.8,0.3,0.0}
\definecolor{DarkCyan}{rgb}{0.0, 0.55, 0.55}

\newcommand{\toolname}{\textsc{Avatar}\xspace}

\newcommand{\find}[1]{
\begin{tcolorbox}[tile,size=fbox,boxsep=1.6mm,boxrule=0pt,top=0pt,bottom=0pt,
borderline west={1mm}{0pt}{blue!50!white},colback=blue!5!white]
\em #1
\end{tcolorbox}
}

\newcommand{\problem}[1]{
\begin{tcolorbox}[tile,size=fbox,boxsep=1.6mm,boxrule=0pt,top=0pt,bottom=0pt,
borderline west={1mm}{0pt}{black!50!white},colback=black!5!white]
\em #1
\end{tcolorbox}
}

\begin{document}

\title{
\toolname: Fixing Semantic Bugs with \\Fix Patterns of Static Analysis Violations
}

\author{
    \IEEEauthorblockN{
        Kui Liu,
        Anil Koyuncu,
        Dongsun Kim,
        Tegawend\'e F. Bissyand\'e
    }
    \IEEEauthorblockA{Interdisciplinary Centre for Security, Reliability and Trust (SnT), University of Luxembourg, Luxembourg
        \\\{kui.liu, anil.koyuncu, dongsun.kim, tegawende.bissyande\}@uni.lu
    }
}

\maketitle

\begin{abstract}
Fix pattern-based patch generation is a promising direction in Automated Program Repair (APR).
Notably, it has been demonstrated to produce more acceptable and correct patches
than the patches obtained with mutation operators through genetic programming. 
The performance of pattern-based APR systems,
however, depends on the fix ingredients mined from fix changes in development histories.
Unfortunately, collecting a reliable set of bug fixes in repositories can be challenging.
In this paper, we propose to investigate the possibility in an APR scenario of leveraging
code changes that address violations by static bug detection tools.
To that end, we build the \toolname APR system, which exploits fix patterns of static
analysis violations as ingredients for patch generation. Evaluated on the Defects4J benchmark,
we show that, assuming a perfect localization of faults, \toolname can generate
correct patches to fix 34/39 bugs. We further find that
\toolname yields performance metrics that are comparable to that of
the closely-related approaches in the literature. While \toolname outperforms
many of the state-of-the-art pattern-based APR systems, it is mostly complementary
to current approaches. Overall, our study highlights the relevance of static
bug finding tools as indirect contributors of fix ingredients for addressing code defects
identified with functional test cases.
\end{abstract}


\vspace{-3mm}
\section{Introduction}
\label{sec:intro}
The current momentum of Automated Program Repair (APR) has led to the development of various approaches in
the literature~\cite{nguyen2013semfix,
westley2009automatically,le2012genprog,kim2013automatic,coker2013program,ke2015repairing,
mechtaev2015directfix,long2015staged,le2016enhancing,le2016history,long2016automatic,chen2017contract,
le2017s3,long2017automatic,xuan2017nopol,xiong2017precise}.
In the software engineering community, the focus is mainly placed on fixing {\em semantic bugs},
i.e., those bugs that make the program behavior deviate from developer's
intention~\cite{mechtaev2018semantic,nguyen2013semfix}. Such bugs are detected by test suites.
APR researchers have then developed repair pipelines where program test cases are
leveraged not only for localizing the bugs but also as the oracle for validating the generated patches.

Unfortunately, given that test suites can be incomplete,
typical APR systems are prone to generate nonsensical patches
that might violate the intended program behavior or simply introduce other defects
which are not covered by the test suites~\cite{kim2013automatic}.
A recent study by Smith et al.~\cite{smith2015cure} has thoroughly investigated this issue
and found that {\em overfitted} patches are actually common: these are patches that can pass all the available test cases, but are not actually {\em correct}.

To address the problem of patch correctness in APR, two research directions are
being investigated in the literature.
The first direction attempts to develop techniques for automatically
augmenting the test suites~\cite{yang2017better}.
The second one focuses on improving the patch generation process to reduce
the probability of generating nonsensical patches~\cite{jiang2018shaping,koyuncu2018fixminer}. The scope of our work is the latter.

Mining fix templates from common patches is a promising approach to achieve patch correctness.
As first introduced by Kim et al.~\cite{kim2013automatic}, patch correctness
can be improved by relying on fix templates learned from human-written patches.
In their work, the template constructions were performed manually,
which is a limiting factor and is further error-prone~\cite{monperrus2014critical}.
Since then, several approaches have been developed
towards automating the inference of fix patterns from fix changes in developer code
bases~\cite{long2017automatic,jiang2018shaping,martinez2014fix,liu2018closer,tufano2018empirical}.
A key challenging step in the inference of patterns, however,
is the identification and collection of a substantial set of relevant
bug fix changes to construct the learning dataset. Patterns must further
be precise and diverse to actually guarantee repair effectiveness.

There have been approaches to mining fix patterns and exploring the challenges in achieving
the diversity and reliability of fix ingredients.
Long et al.~\cite{long2017automatic}
have relied on only three simple bug types, while Koyuncu et al.~\cite{koyuncu2018fixminer}
have focused on bug linking between bug tracking systems and
source code management systems to identify probable bug fixes.
Unfortunately, the former approach cannot find patterns to address a variety of bugs,
while the latter may include patterns that are irrelevant to bug fixes
since developer changes are not atomic~\cite{herzig2013impact};
it is thus challenging to extract useful and reliable patterns focusing on fix changes.

Our work proposes a new direction for pattern-based APR to overcome
the limitations in finding reliable and diverse fix ingredients. Concretely,
we focus on developer patches that are fixing static analysis violations.
The advantages of this approach are: (1) the availability of toolsets
for assessing whether a code change is actually a fix~\cite{heckman2008establishing,avgustinov2015tracking},
and (2) the ability to further pre-categorize
the changes into groups targeting specific violations,
leading to consistent fix patterns~\cite{liu2019mining,rolim2018learning}.
Although static analysis violations (e.g., FindBugs~\cite{findbugs} warnings) may
appear irrelevant to the problem of {\em semantic} bug fixing,
there are two findings in the literature, which can support our intuition of leveraging
fix patterns from static analysis violation patches to address semantic bugs:
\begin{itemize}[leftmargin=*]
	\item {\em Locations of {\bf\em semantic bugs} (unveiled through dynamic execution of test cases) can sometimes be detected by static analysis tools.}
    In a recent study, Habib et al.~\cite{habib2018many}, have found that some bugs in the Defects4J
    dataset can be identified by static analysis tools: SpotBugs~\cite{spotbugs}, Infer~\cite{fbInfer} and
    ErrorProne~\cite{errorprone}. Other studies~\cite{fontana2011experience,moha2010decor,yamashita2013developers} have also suggested
    that violations reported by static analysis tools might be smells of more severe defects
    in software programs.
	\item {\em Violation fix patterns have been used to successfully fix bugs in the wild.}
	In preliminary live studies, Liu et al.~\cite{liu2019mining} and Rolim et al.~\cite{rolim2018learning} have shown that it can systematically fix statically detected bugs by using some of their previously-learned fix patterns. They further showed that project developers are even eager to integrate the systematization of such fixes based on the mined patterns.
\end{itemize}

\problem{Our work investigates to what extent fix patterns for static analysis violations can serve as ingredients to the patch generation step in an automated program repair pipeline.}

This paper thus makes the following contributions:
\begin{enumerate}[leftmargin=*]
	\item {\em We discuss a counterpoint to a recent study in the literature on the usefulness
of static analysis tools to address real bugs}. We find that, although static bug finding tools, can detect a relatively small number of real-world semantic bugs from the Defects4J dataset, fix patterns inferred from the patches addressing statically detected bugs can provide relevant ingredients in an APR pipeline that is targeting semantic bugs.
	\item {\em We propose} \toolname (static \underline{a}nalysis \underline{v}iol\underline{a}tion fix pa\underline{t}tern-based \underline{a}utomated program \underline{r}epair), 
{\em a novel fix pattern-based approach to automated program repair.}
   Our approach differs from related work in the dataset of developer patches that is leveraged to extract fix ingredients.
    We build on patterns extracted from patches that have been verified (with bug detection tools) as true bug fix patches. Given the redundancy of bug types detected by static analysis tools, the associated fixes are intuitively more similar, leading to the inference of reliable common fix patterns.
    \toolname is available at:
    {\url{https://github.com/SerVal-Repair/AVATAR}}.
	\item {\em We report on an empirical assessment of \toolname on the Defects4J benchmark.} We compare our approach with the state-of-the-art based on different evaluation aspects,
    including the number of fixed bugs, the exclusivity of fixed bugs, patch correctness, etc.
    Among several findings, we find that \toolname is capable of generating correct patches for 34 bugs, and partially-correct patches for 5 bugs, when assuming a perfect fault localization scenario.
\end{enumerate}

\vspace{-1mm}
\section{Background}
\label{sec:bg}
We provide background information on general pattern-based APR, as well as on pattern inference from static analysis violation data.

\vspace{-2mm}
\subsection{Automated Program Repair with Fix Patterns}
\label{sec:fixpattern:mining}
Fix pattern-based APR has been widely
explored in the
literature~\cite{kim2013automatic,le2016history,long2016automatic,tan2016anti,long2017automatic,
saha2017elixir}. The basic idea is to abstract a code change into a {\em pattern} (interchangeably referred to as a {\em template}) that can be applied to a faulty code. The fixing process thus consists in leveraging context information of faulty code (e.g., abstract syntax tree (AST) nodes)
to match context constraints defined in a given fix pattern. For example, the fix template ``Method Replacer''
provided in PAR~\cite{kim2013automatic} is presented as:
\begin{center}
	{\tt  obj.{\bf\ttfamily method1}(param)} $\rightarrow$ \ttfamily{obj.{\bf\ttfamily method2}(param)}
\end{center}
where the faulty method call {\tt method1} is replaced by another method call
{\tt method2} with compatible parameters and return type.
A method call is the context information for this template to match buggy code fragment.
Thus, this template can be applied to any faulty statement that includes at least one method call expression. The template further
guides the patch candidate generation where changes are proposed to replace the potentially faulty method call with another method call.

Mining fix patterns has some intrinsic issues.
The first issue relates to the variety of patterns that must be identified to support the fixing of different bug types.
There are two strategies in fix pattern mining: (1) manual design and (2) automatic mining.
The former can effectively create precise fix patterns. Unfortunately, it  requires
human effort, which can be prohibitive~\cite{kim2013automatic}. The latter
infers common modification rules~\cite{long2017automatic} or searches for the most redundant sub-patch
instance~\cite{koyuncu2018fixminer,jiang2018shaping}.
While this strategy can substantially increase the number of fix patterns, it is subject to noisy input data due to
{\em tangled changes}~\cite{herzig2013impact}, which make the inferred patterns less relevant.
The second issue relates to the granularity (i.e., the degree of abstraction). Coarse-grained and monolithic patterns~\cite{pan2009toward}
can cover many types of bugs but they may not be actionable in APR.
A fine-grained or micro pattern~\cite{long2017automatic} can be readily actionable, but
cannot cover many defects.

\vspace{-2mm}
\subsection{Static Analysis Violations}

Static analysis tools help developers check for common programming errors in software systems.
The targeted errors include syntactic defects, security vulnerabilities, performance issues, and
bad programming practices. 
These tools are qualified as ``static'' because they do not require dynamic execution traces to find bugs. Instead, they are directly applied to source code or bytecode. In contrast with dynamic analysis tools which must run test cases, static tools can cover more paths, although it makes over-approximations that make them prone to false positives.

Many software projects rigorously integrate static analysis tools into their development cycles. The Linux kernel development project is such an example project where developers systematically run static analyzers against their code before pushing it to maintainers repositories~\cite{koyuncu2017impact}. 
More generally, FindBugs~\cite{findbugs}, PMD~\cite{pmdtool} and Google Error-Prone~\cite{errorprone}
are often used in Java projects, while C/C++ projects
tend to adopt Splint~\cite{splint}, cppcheck~\cite{cppcheck},
and Clang Static Analyzer~\cite{clangsa}. 

\begin{figure*}[!t]
	\centering
    \includegraphics[width=.85\linewidth]{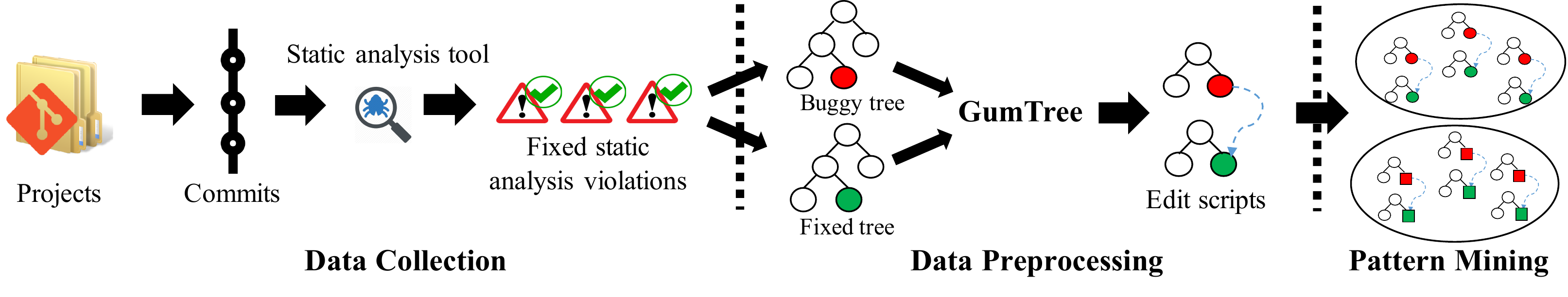}
    \vspace{-4mm}
    \caption{Summarized steps of static analysis violation fix pattern mining.}
    \label{fig:mingingProcess}
\end{figure*}

Static analysis tools raise {warnings}, which are also referred to as {alerts}, {alarms}, or violations. Given that these warnings are due to the detection of code fragments that do not comply with some analysis rules, in the remainder of this paper we refer to the issues reported by static analysis tools as {\em violations}.

Figure~\ref{fig:actionable:violation} shows an example
patch for a violation detected by FindBugs.
This violation (the \textcolor{red}{red} code) is reported because the {\tt equals} method should work for all object types (i.e., {\tt Object}): 
in this case, the method code violates the rule since it assumes a specific type (i.e., {\tt ModuleWrapper}).

\begin{figure}[!tp]
\begin{lstlisting}[language=diff,linewidth={\linewidth},frame=tb,basicstyle=\scriptsize\ttfamily]
 public boolean equals(Object obj) {
 // (@{\bf Violation Type:}@) (@{\tt\fontsize{6.8}{6.8}\selectfont BC\_EQUALS\_METHOD\_SHOULD\_WORK\_FOR\_ALL\_OBJECTS}@)
-    return getModule().equals(}
-           ((ModuleWrapper) obj).getModule());
+    return obj instanceof ModuleWrapper &&
+           getModule().equals(
+           ((ModuleWrapper) obj).getModule());
 }
\end{lstlisting}
    \vspace{-1.5mm}
    \caption{Example patch for fixing a violation detected by FindBugs. Example excerpted from~\cite{liu2019mining}.}
    \label{fig:actionable:violation} 

\end{figure}

Note that, not all violations are accepted by developers as actual defects. Since static analysis tools
use limited information, detected violations could be correct code (i.e., false positive)
or the warning may be irrelevant (e.g., cannot occur at runtime, or not a serious issue).
In the literature, many studies assume that a violation can be classified as
actionable if it is discarded after a developer changed the location where the violation is detected.
The violation in Figure~\ref{fig:actionable:violation}~\cite{liu2019mining} is fixed by adding an
{\tt instanceof} check (c.f., the \textcolor{codegreen}{green} code in the patch diff); this violation can thus be regarded as {\em actionable} since
this violation is gone after fixing its source code.

\paragraph*{{\bf Motivation}}
Mining patterns from developer patches that fix static analysis violations may help overcome the issues of
fix pattern mining described in Section~\ref{sec:fixpattern:mining}.
First, since static analysis tools specify the type of each violation (e.g., bug descriptions\footnote{\url{http://findbugs.sourceforge.net/bugDescriptions.html}} of FindBugs), each bug instance is already classified as long as it is
fixed by code changes. Thus, we can reduce the manual effort to
collect and classify bugs and their corresponding patches for fix pattern mining.
Second, we can mitigate the issue of tangled changes~\cite{herzig2013impact}
because violation-fixing changes can be localized by static analysis tools~\cite{avgustinov2015tracking}.
Finally, the granularity of fix patterns can be appropriately adjusted for each violation type
since static analysis tools often provide  information on the scope  of
each violation instance.

\section{Mining Fix Patterns for Static Violations}

Mining fix patterns for static analysis violations has recently been explored in the
literature~\cite{liu2019mining,rolim2018learning}. The general objective so far, however,
is to learn quick fixes for speeding maintenance tasks and towards understanding
which violations are prioritized by developers for fixing.
To the best of our knowledge, our work is the first reported attempt to investigate fix patterns of
static analysis violations in the context of automated program repair
(where patches are generated and validated systematically with developer test cases).

There have been two recent studies of mining fix patterns addressing static analysis
violations.
Our previous study~\cite{liu2019mining} focuses on identifying fix
patterns for FindBugs violations~\cite{hovemeyer2004finding},
while Rolim et al.~\cite{rolim2018learning} consider PMD violations~\cite{copeland2005pmd}.
Both approaches, which were developed concurrently, leverage a similar methodology
in the inference process.
We summarise below the process of fix pattern mining of static analysis violations into three basic steps
(as shown in Figure~\ref{fig:mingingProcess}):
data collection, data preprocessing, and fix pattern mining.
Implementation details are strictly based on the approach of our previous study~\cite{liu2019mining}.

\begin{figure*}[!t]
	\centering
    \includegraphics[width=.85\linewidth]{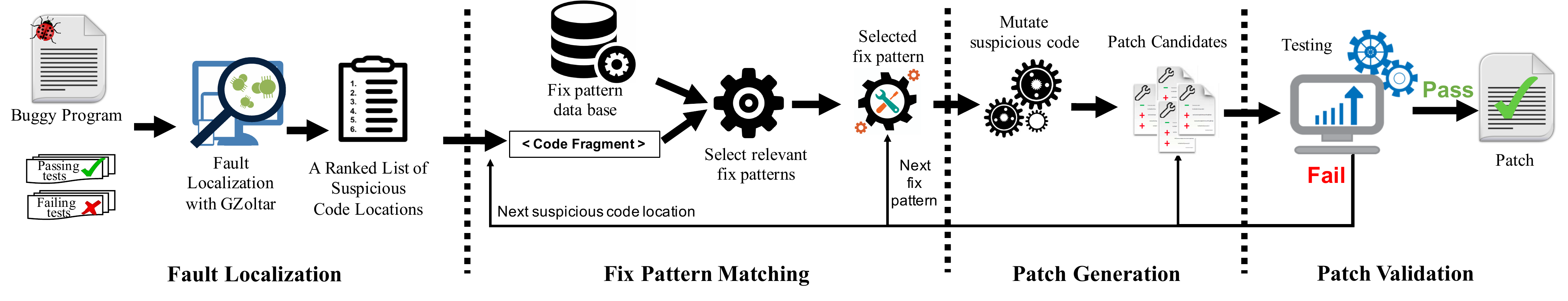}
\vspace{-3mm}
    \caption{Overview bug fixing process with \toolname.}
    \label{fig:aprApproach}
\end{figure*}

\vspace{-2mm}
\subsection{Data Collection}
The objective of this step is to collect patches that are relevant to static analysis violations.
This step is done in the wild based on the commit history of open-source projects
by implementing a strict strategy to limit the dataset of changes to those that are relevant
in the context of static analysis violations.
To that end, it is necessary to systematically run static bug detection tools to each and
every revision of the programs.
This process can be resource-intensive: for example, FindBugs takes as input
compiled versions of Java classes, requiring to build thousands of project revisions.

This step collects code changes (i.e., patches) only if they are identified as
violation-fixing changes.
For a given violation instance, we can assume that a change commit is a (candidate) fix for the instance
when it disappears after the commit:
i.e., the violation instance is identified in a revision of a program,
but is no longer identified in the consecutive revision.
Then, it is necessary to figure out whether
the change actually fixed the violation instance or
it just disappears by coincidence.
If the affected code lines are located within the code change
diff
\footnote{A ``code change diff'' consists of two code snapshots. One snapshot
represents the code fragment that will be affected by a code change, while
the other one represents the code fragment after it has been affected by
the code change.}
of the commit, it is regarded as an actual fix for the given violation instance.
Otherwise, the violation instance might be removed just by
deleting a method, class, or even a file.
Eventually, all code change diffs associated to the identified fixed violation instances are collected
to form the input data for fix pattern mining.

\vspace{-1mm}
\subsection{Data Preprocessing}
Once violation patch data are collected,
they are processed to extract concrete change actions.
Patches submitted to program repositories are presented in the form of line-based GNU diffs
where changes are reported in a text-based format of edit script.
Given that, in modern programming languages, such as Java, source code lines
do not represent a semantic entity of a code entity
(e.g., a statement may span across several lines),
it is challenging to directly mine fix patterns from GNU diffs.

Pattern-mining studies leverage edit scripts of program
Abstract Syntax Trees (ASTs). Concretely, the buggy version
(i.e., program revision file where the violation can be found)
and the fixed version (i.e., consecutive program revision file where the violation does not appear)
are given as inputs to the GumTree~\cite{falleri2014fine}, an AST-based code differencing tool, to
produce the relevant AST edit script.
This edit script describes in a fine-grained manner the repair actions that are
implemented in the patch. Figure~\ref{fig:bugCl31} provides an example GNU Diff for a bug fix patch, while Figure~\ref{fig:editbugCl31} illustrates the associated AST edit scripts.

\begin{figure}[!h]
    \centering
\begin{lstlisting}[language=diff,linewidth={\linewidth},frame=tb,basicstyle=\scriptsize\ttfamily]
--- a/src/com/google/javascript/jscomp/Compiler.java
+++ b/src/com/google/javascript/jscomp/Compiler.java
@@ -1283,4 +1283,3 @@
	// Check if the sources need to be re-ordered.            
	if (options.dependencyOptions.needsManagement() &&
-              !options.skipAllPasses &&
               options.closurePass) {

// (@{\bf\ttfamily Defects4J Dissection: }@)
// Repair Action: Conditional expression reduction.
\end{lstlisting}
    \vspace{-1.5mm}
    \caption{Patch of the bug \protect{Closure-13}\protect\footnotemark{} in Defects4J.}
    \label{fig:bugCl31}
\end{figure}
\footnotetext{\url{http://program-repair.org/defects4j-dissection/#!/bug/Closure/31}}

\begin{figure}[!h]
	\centering
\begin{lstlisting}[linewidth={\linewidth},frame=tb,basicstyle=\scriptsize\ttfamily]
UPD IfStatement@@(@{``if statement code''}@)
---UPD InfixExpression@@(@{``infix-expression code''}@)
------DEL PrefixExpression@@(@{``!options.skipAllPasses''}@)
------DEL Operator@@(@{``\&\&''}@)
\end{lstlisting}
    \vspace{-1.5mm}
	\caption{AST edit scripts produced by GumTree for the patch in Fig.~\ref{fig:bugCl31}.}
	\label{fig:editbugCl31}
\end{figure}

\vspace{-1mm}
\subsection{Fix Pattern Mining}
Given a set of edit scripts, the objective of the pattern mining step is to group
``similar'' scripts in order to infer the common subset of edit actions,
i.e., a consistent pattern across the group.
To that end, Rolim et al.~\cite{rolim2018learning} rely on the greedy algorithm to compute the
distance among edit scripts. Edit scripts with low distances among them are grouped together.
Our previous study~\cite{liu2019mining}, on the other hand, leverage
a deep representation learning framework (namely, CNNs~\cite{matsugu2003subject}) to
learn features of edit scripts, which are then used to find clusters of similar edit scripts.
Clustering is performed based on the X-means algorithm.
Finally, the largest common subset of edit actions among all edit scripts in a cluster
is considered as the pattern.

Mined fix patterns with this approach have already been proven useful by the authors.
For example, our previous study~\cite{liu2019mining} and Rolim's work~\cite{rolim2018learning} conducted
live studies by making pull requests to projects in the wild:
the pull requests contained change details of a patch that is generated
based on the inferred fix patterns to fix static analysis violations in developer code.
Developers accepted to merge 67 out of 116 patches generated for FindBugs violations in our previous study~\cite{liu2019mining}.
Similarly, 6 out of 16 pull requests by Rolim et al.~\cite{rolim2018learning} have been merged by developers in the wild.
Such promising results demonstrated the possibility to automatically fix bugs that are addressed by static bug detection tools.

\find{Fix patterns of static analysis violations have been explored in the literature
to automate patch generation for bugs that are statically detected.
To the best of our knowledge, \toolname is the first attempt to leverage fine-grained  patterns
of static analysis violations as fix ingredients for automated program repair that addresses
semantic bugs revealed by test cases.}

\vspace{-2mm}
\section{Our Approach}
\label{sec:method}

As shown in Figure~\ref{fig:aprApproach}, \toolname consists of four major
steps for automated program repair. In this section, we detail the objective
and design of each step, and provide concrete information
on implementation.

\vspace{-1mm}
\subsection{Fault Localization}
We rely on the GZoltar\footnote{\url{http://www.gzoltar.com}}~\cite{campos2012gzoltar}
framework to automate the execution of test cases for each program.
In the framework, we leverage the Ochiai~\cite{abreu2007accuracy} ranking metric to actually compute
the suspiciousness scores of statements that are likely to be the faulty code locations.
This ranking metric has been demonstrated in several empirical
studies~\cite{steimann2013threats,xie2013theoretical,xuan2014learning,pearson2017evaluating}
to be effective for localizing faults in object-oriented programs.
The GZoltar framework for fault localization is also widely used in the literature of
APR~\cite{martinez2016astor,xiong2017precise,xuan2017nopol,xin2017leveraging,wen2018context,koyuncu2018fixminer,kui2018live,jiang2018shaping}.

\vspace{-1mm}
\subsection{Fix Pattern Matching}
In the running of the repair pipeline, once fault localization produces a list of suspicious code locations,
\toolname iteratively attempts to match each of these locations with a given pattern from the database of mined fix patterns.
Fix patterns in our database are collected from the artifacts released by Liu et al.~\cite{liu2019mining}
and Rolim et al.~\cite{rolim2018learning}.
Table~\ref{tab:patterns} shows statistics about the pattern collection in these previous works. As most of the fix patterns released by Liu et al. will not change the program behavior, we only select 13 of them for 10 violation types after manually checking that they can change the program behavior (details shown in the aforementioned website).
\begin{table}[!tp]
	\caption{Statistics on fix patterns of static analysis violations.}
	\vspace{-1.5mm}
	\begin{tabular}{lcccc}
		& \# Projects & \# violation  & \# violation & \# fix \\
		& & fix patches &  types & patterns \\
	\toprule
	Liu et al.~\cite{liu2019mining}	& 730 & 88,927 & 111 & 174 \\ \hline
	Rolim et al.~\cite{rolim2018learning} & 9 &  288,899 & 9 & 9\\
	\bottomrule
	\end{tabular}
	\label{tab:patterns}
\end{table}

Recall that each pattern is actually an edit script of repair actions on specific AST node types.
AST nodes associated to the faulty code locations are then regarded as the {\em context}
of matching the fixing patterns: i.e., these nodes are checked against the nodes involved
in the edit scripts of fix patterns.
For example, the fix pattern shown in Figure~\ref{fig:uselessCond}
contains three levels of contexts:
(1) {\tt
IfStatement}, 
which means that the pattern
is matched only if the suspicious faulty statement is an {\tt IfStatement};
(2) {\tt InfixExpression} 
indicates that the pattern is relevant when the predicate expression of the suspicious {\tt IfStatement} is an {\tt InfixExpression};
(3) the matched {\tt InfixExpression} predicate in the suspicious statement must contain at least two sub-predicate expressions.

\begin{figure}[!h]
	\centering
\begin{lstlisting}[linewidth={\linewidth},frame=tb,basicstyle=\scriptsize\ttfamily]
// (@{\bf\ttfamily Fix Pattern: Remove a useless sub-predicate expression.}@)
UPD IfStatement
---UPD InfixExpression@expA Operator expB
------DEL Expression@expB
------DEL Operator
\end{lstlisting}
    \vspace{-1.5mm}
	\caption{A fix pattern for \protect{\tt UC\_USELESS\_CONDITION}\protect\footnotemark{} violation~\cite{liu2019mining}.}
	\label{fig:uselessCond}
\end{figure}

\addtocounter{footnote}{-1}
\stepcounter{footnote}\footnotetext{The condition has no effect and always produces the same result as the value of the involved variable was narrowed before. Probably something else was meant or condition can be removed.}

A pattern is found to be relevant to a faulty code location only if all AST node contexts at this location
matches with the AST node of the pattern.
For example, the bug shown in Figure~\ref{fig:bugCl31} is located within an {\tt IfStatement} with an {\tt InfixExpression}
which is formed by three sub-predicate expressions. This buggy fragment thus matches
the fix pattern shown in Figure~\ref{fig:uselessCond}.

\vspace{-1mm}
\subsection{Patch Generation}
Given a suspicious code location and an associated matching fix pattern, \toolname applies the repair
actions in the edit scripts of the pattern to generate patch candidates.
For example, the code change action of the fix pattern in Figure~\ref{fig:uselessCond} is interpreted as
removing a sub-condition expression (or sub-predicate expression) in a faulty {\tt IfStatement}. Thus,
three patch candidates, as shown in Figure~\ref{fig:pCan}, can be generated by \toolname for the buggy code
in Figure~\ref{fig:bugCl31} since the statement has \uline{three} candidate sub-predicates expressions.

\vspace{-1mm}
\subsection{Patch Validation}
\label{sec:patchValidation}
Patch candidates generated by \toolname must then be systematically assessed.
Eventually, using test cases, our approach verifies whether a patch candidate is
a {\em plausible} patch or not. We target two types of plausible patches:

\begin{itemize}
	\item {\bf \em Fully-fixing} patches, which are patches that make the program
	pass all available test cases. Once such a patch is validated, the execution
	iterations of \toolname are halted.
	\item {\bf \em Partially-fixing} patches, which are patches that make the program
	pass not only all previously-passing test cases, but also part of the previously-failing test cases.
\end{itemize}

The first generated {\em fully-fixing} patch is prioritized over
any other generated patch, and is considered as the plausible patch for the given bug.
After iterating over all suspicious statements with all matching fix patterns,
if \toolname fails to generate a {\em fully-fixing} patch for a bug,
but generates some {\em partially-fixing} patches,
these patches are considered as plausible patches.
Nevertheless, we implement a selection scheme where {\em partially-fixing} patches
that change the program control-flow are prioritized over those that only change data-flow.

{\em Partially-fixing} patches that change the control flow are further na\"ively
ordered by the edit distances (at AST node level) between the patched
fragments and the buggy fragment: smaller edit distances are preferred.
Ties are broken by considering precedence in the generation:
the first generated {\em partially-fixing} patch is the final plausible patch.

\begin{figure}[!tp]
    \centering
    \vspace{1mm}
\begin{lstlisting}[language=diff,linewidth={\linewidth},frame=tb,basicstyle=\scriptsize\ttfamily]
  // (@{\bf\ttfamily Patch Candidate I.}@)
- if (options.dependencyOptions.needsManagement() &&
-              !options.skipAllPasses &&
+ if (!options.skipAllPasses &&
               options.closurePass) {

  // (@{\bf\ttfamily Patch Candidate II.}@)
  if (options.dependencyOptions.needsManagement() &&
-              !options.skipAllPasses &&
               options.closurePass) {

  // (@{\bf\ttfamily Patch Candidate III.}@)
  if (options.dependencyOptions.needsManagement() &&
-              !options.skipAllPasses &&
-               options.closurePass) {
+              !options.skipAllPasses) {
\end{lstlisting}\vspace{-1.5mm}
    \caption{Patch Candidates generated by \toolname with a fix pattern that is mined from patches for {\tt UC\_USELESS\_CONDITION} violations (cf. Fig.~\ref{fig:uselessCond}), and which matches the buggy statement in  Closure-13 bug (cf. Fig.~\ref{fig:bugCl31}).}
    \label{fig:pCan}
\end{figure}

\section{Assessment}
\label{sec:eval}

We evaluate \toolname on Defects4J~\cite{just2014defects4j}, which is widely used by state-of-the-art
APR systems targeting Java programs.
Table~\ref{tab:defects} summarizes the statistics on the number of bugs and
test cases available in the version 1.2.0\footnote{\url{https://github.com/rjust/defects4j/releases/tag/v1.2.0}}
of Defects4J. 

\begin{table}[!h]
	\centering
	\caption{Defects4J dataset information.}
	\vspace{-1mm}
	{\scriptsize
	\begin{threeparttable}
		\begin{tabular}{l|ccc}
			\toprule
			{\bf Project} & {\bf Bugs} & {\bf kLoC} & {\bf Tests} \\
			\hline
			JFreeChart (Chart)         & 26  & 96   & 2,205\\
			Closure compiler (Closure) & 133 & 90   & 7,927\\
			Apache commons-lang (Lang) & 65  & 22   & 2,245\\
			Apache commons-math (Math) & 106 & 85   & 3,602\\
			Mockito                    & 38  & 11   & 1,457\\
			Joda-Time (Time)           & 27  & 28   & 4,130\\
			\hline
			Total                      & 395 & 332 & 21,566 \\
			\bottomrule
		\end{tabular}
	\end{threeparttable}
	}
	\label{tab:defects}

\end{table}

``{\bf Bugs}'', ``{\bf kLoC}'', and ``{\bf Tests}'' denote respectively the number of bugs, the program size in kLoC (i.e., thousands of lines of code), and the number of test cases for each subject.
The overall number of kLoC and test cases for project Mockito are not indicated in the Defects4J
paper~\cite{just2014defects4j} from where the reported information is excerpted.

\vspace{-1mm}
\subsection{Research Questions}

Our investigation into the repair performance of \toolname seeks to answer
the following research questions (RQs):
\begin{itemize}[leftmargin=*]
	\item {\bf RQ1:} {\bf\em How effective are fix patterns of static analysis violations for repairing
	programs with semantic bugs?} Recall that we broadly consider as {\em semantic bugs} all bugs that are uncovered by executing developer {\em test cases}. Our first research question assesses how many bugs in the Defects4J benchmark
	can be fixed with fix patterns of static analysis violations.
	To that end, we first (1) investigate how
	effectively \toolname can fix such semantic bugs that appear to be localizable by static analysis tools.
	Then, (2) building on the assumption that the location of the semantic bug is known,
	we investigate whether \toolname can generate a correct patch to fix it.
	\item {\bf RQ2:} {\bf\em Which bugs and which patterns are relevant targets for \toolname in an automated
	program repair scenario?} This research question dissects the data yielded during the investigation of
	RQ1, with the objective of assessing the diversity of bugs that can be fixed as well as the types of
	violation fix patterns that have been successfully leveraged.
	\item {\bf RQ3:} {\bf\em How does \toolname compare to the state-of-the-art with respect to repair
	performance?} With this research question, we aim at showing whether the proposed approach is relevant in
	the landscape of APR systems. Does \toolname offer comparable performance? To what extent can \toolname
	complement existing APR systems?
\end{itemize}

\vspace{-2mm}
\subsection{Experimental Setup}

For evaluation purpose, we apply different fault localization schemes to the experiment
of each RQ, while the default setting of \toolname is to use the GZoltar framework with the Ochiai ranking metric for ordering suspicious statements. The usage of GZoltar and Ochiai reduces the comparison biases since both are widely used by APR systems in the literature.
\begin{itemize}[leftmargin=*]
	\item First, we apply \toolname to Defects4J bugs that are localized by three state-of-the-art static
	analysis tools (namely, SpotBugs~\cite{spotbugs}, Facebook Infer\cite{fbInfer}, and Google ErrorProne~\cite{errorprone}) (for RQ1; see Section~\ref{sec:rq1:localizable}). To that end, we consider recent
	data reported by Habib and Pradel~\cite{habib2018many}. This configuration focuses on the effectiveness
	of \toolname on \uline{such semantic bugs that can also be detected statically}.
	\item Second, we apply \toolname on all faulty code positions in the benchmark (for RQ1; see Section~\ref{sec:frb}). We thus
	\uline{assume that a perfect localization is possible}, and assess the performance
	of the approach on all bugs.
	\item Finally, for RQ3, we compare \toolname with the state-of-the-art APR tools that are evaluated on
	the Defects4J benchmark (see Section~\ref{sec:comparison}). To that end, we attempt to replicate two scenarios of fault localization used in
	APR assessments: the first scenario assumes that \uline{the faulty method name is
	known}~\cite{le2016history} and thus focuses on ranking the inner-statements based on Ochiai
	suspiciousness scores; the second scenario \uline{makes no assumption on fault location} and thus uses the default setting of \toolname.
\end{itemize}

\vspace{-2mm}
\subsection{Applying \toolname to Statically-Detected Bugs in Defects4J}
\label{sec:rq1:localizable}
Table~\ref{tab:saBugs} provides details on the Defects4J bugs that can be detected by static analysis tools
and are successfully repaired by \toolname.
We report that out of the 14, 4, and 7
bugs that can be detected respectively by SpotBugs, Facebook Infer and Google ErrorProne on version 1.2.0
of Defects4J, \toolname can successfully generate correct patches for 3, 2 and 1 bugs.

Overall four distinct localizable bugs have been correctly fixed with patches generated from the fix
patterns that were mined from patches fixing FindBugs violations~\cite{liu2019mining}.
All the four bugs are related to distinct violation types.
In their work, Liu et al.~\cite{liu2019mining} claimed that their mined patterns could be applied to
violations reported by other static analysis tools.
Our experiment indeed shows that these patterns fixed two
bugs detected by SpotBugs (i.e., the successor of FindBugs), which are also detected by Facebook Infer.

\begin{table}[!tp]
	\centering
	\caption{Statically-detected bugs fixed by \toolname.}
	\vspace{-1.5mm}
	\label{tab:saBugs}
{\scriptsize
\setlength{\tabcolsep}{1pt}
	\begin{threeparttable}
		\begin{tabular}{L{12mm}|ccc|c}
			\toprule
			{\bf Bug ID} & {\bf SpotBugs} & {\bf  Infer} & {\bf  ErrorProne} & {\bf Static Analysis Violation Type} \\
			\hline
			Chart-1      & \ding{108} & \ding{108} &      & \protect{\tt NP\_ALWAYS\_NULL}\protect\footnotemark{}\\
			Chart-4      & \ding{108} & \ding{108} & & \protect{\tt NP\_NULL\_ON\_SOME\_PATH}\protect\footnotemark{} \\
			Chart-24     & \ding{108} &            & & \protect{\tt DLS\_DEAD\_LOCAL\_STORE}\protect\footnotemark{} \\
			Math-77      &  & & \ding{108} &  \protect{\tt UPM\_UNCALLED\_PRIVATE\_METHOD}\protect\footnotemark{}\\
			\hline
			Total   & 3/14 (18) & 2/4 (5) & 1/7 (8) &  \\
			\bottomrule
		\end{tabular}
		{\footnotesize $x/y(z)$ reads as: $x$ is the number of bugs fixed by \toolname
among the $y$ bugs in version 1.2.0 of Defects4J that can be localized by each static analysis tool. $z$ is
given as an indicator for the number of statically localizable bugs in an augmented
version\protect\footnotemark  of Defects4J. The information on
localizable bugs is excerpted from the study of Habib and Pradel~\cite{habib2018many}. Since most of
the state-of-the-art APR systems targeting Java program are evaluated
on the version 1.2.0, our experiments focus on localizable bugs in this version.
}
	\end{threeparttable}
	}
\end{table}

\begin{table*}[!t]
	\centering
	\caption{Number of Defects4J bugs fixed by \toolname with an assumption of perfect localization information.}
	\vspace{-1.5mm}
	\label{tab:fixedBugs}
	\begin{threeparttable}
		\begin{tabular}{l|cccccc|C{10mm}}
			\toprule
			{\bf Fixed Bugs} & {\bf Chart (C)} &  {\bf Closure (Cl)} & {\bf Lang (L)} & {\bf Math (M)} & {\bf Mockito (Moc)} & {\bf Time (T)} & {\bf Total} \\ \hline
			\# of Fully Fixed Bugs & 7/8   & 10/13 & 5/10  & 8/13 & 2/2 & 2/3 & 34/49\\
			\# Partially Fixed Bugs & 2/3  & 1/4  & 1/3  & 1/4   & 0/0  & 0/0 & 5/14\\
			\bottomrule
		\end{tabular}
		{\footnotesize $^\dagger$ In each column, we provide $x/y$ numbers: $x$ is the number of correctly fixed bugs; $y$ is the number of bugs for which a plausible patch is generated by the APR tool. The same as the following similar tables.}
	\end{threeparttable}
\end{table*}

\addtocounter{footnote}{-5}
\stepcounter{footnote}\footnotetext{A null pointer is dereferenced and will
lead to a NullPointerException when the code is executed.}
\stepcounter{footnote}\footnotetext{Possible null pointer dereference.}
\stepcounter{footnote}\footnotetext{Dead store to a local variable.}
\stepcounter{footnote}\footnotetext{A private method is never called.}
\stepcounter{footnote}\footnotetext{\url{https://github.com/rjust/defects4j/pull/112}}

\find{{\bf RQ1$\blacktriangleright$}
\toolname demonstrates the usefulness of violation fix patterns
in a scenario of automating some maintenance tasks
involving systematic checking of developer code with static checkers.}

	Our experiments, however, reveal that \toolname's fix patterns for static analysis violations are not effective on many \uline{supposedly} statically-detectable bugs in Defects4J. We investigate the cases of such bugs, and find that:
\begin{enumerate}
	\item some of the bugs have a code context that does not match any of the mined fix patterns;
	\item in some cases, the detection is actually a coincidental false positive reported in~\cite{habib2018many}, since the violation does not really match the semantically faulty code entity that must be modified. Figure~\ref{fig:falsePositive} provides a descriptive example of such false positives.
	\item finally, in other cases, the fixes are truly domain-specific, and no pattern is thus applicable.
\end{enumerate}

\begin{figure}[!h]
    \centering
\begin{lstlisting}[language=diff,linewidth={\linewidth},frame=tb,basicstyle=\scriptsize\ttfamily]
// (@{\bf\ttfamily Violation Type: }@)FE_FLOATING_POINT_EQUALITY
// The comparing result of two floating point values for 
// equality may not be accurate.

// (@{\bf\ttfamily Defects4J Dissection: }@)
// Bug Pattern: Conditional block removal.

- if (x == x1) {
-   x0=0.5*(x0+x1-FastMath.max(rtol*FastMath.abs(x1),atol));
-   f0=computeObjectiveValue(x0);
- }
  break; 
\end{lstlisting}
    \vspace{-1.5mm}
    \caption{A bug is false-positively identified as a statically-detected bug in~\cite{habib2018many} (Math-50): the violation is not related to the test case failures.}
    \label{fig:falsePositive}
\end{figure}

\vspace{-2mm}
\subsection{Applying \toolname to All Defects4J Bugs}
\label{sec:frb}
After focusing on those Defects4J bugs that can be statically detected, we run \toolname on all the dataset bugs. Given that the objective is to assess whether a correct patch can be generated if the bug is known, we assume in this experiment that the faulty code locations are known. Concretely, we do not rely on any fault localization tool. Instead, we consider the ground truth of developer patches and list the locations that have been modified as the faulty locations.

The repair operations thus consist in generating patches for the relevant bug locations.
Table~\ref{tab:fixedBugs} details the number of Defects4J bugs that are fixed by \toolname.
Fully and partially fixed bugs are fixed with {\em fully-fixing} and {\em partially-fixing} patches (c.f. Section~\ref{sec:patchValidation}) generated by \toolname, respectively. 
Overall, \toolname can fix 49 bugs with plausible patches (i.e., that pass all available tests).
35 of them are further manually confirmed to be {\em correct} 
(i.e., they are syntactically or at least semantically equivalent to the ground truth patches submitted by developers).
We also note that, for 14 other bugs, \toolname generates plausible patches that make the program pass some previously-failing test cases, without failing any of the previously-passing test cases.
Five among these patches are manually found to be correctly fixing part of the bugs. To the best of our knowledge, \toolname is the first APR tool which partially, but correctly, fixes bugs in Defects4J that have multiple fault code locations.

\find{{\bf RQ1$\blacktriangleright$}\toolname can effectively fix several semantic bugs from the Defects4J dataset. We even observe that the fine-grained fix ingredients can be helpful to target bugs with multiple faulty code locations.}

\begin{table*}[!t]
	\centering
	\caption{Fix ingredients leveraged in the static analysis violation fix patterns used by \toolname to correctly fix semantic bugs.}
	\vspace{-1.5mm}
	\label{tab:fixedBugsFP}
	{\scriptsize
	\begin{threeparttable}
		\begin{tabular}{l|l|c}
			\toprule
			{\bf Violation Types associated with the Fix Patterns} & {\bf Fix Ingredients from the Violation Fix Patterns}&  {\bf Fixed Bug IDs} \\
			\hline\hline
			{\tt NP\_ALWAYS\_NULL}               & \makecell[l]{Mutate the operator of null-check expression:\\
													``{\tt  var \textcolor{red}{\sout{\ttfamily\bf!=} }\textcolor{codegreen}{==} null}'', or
													``{\tt var \textcolor{red}{\sout{==} }\textcolor{codegreen}{!=} null}''.}& C-1.\\\midrule
			{\tt NP\_NULL\_ON\_SOME\_PATH}       & \makecell[l]{
													1. Wrap buggy code with if-non-null-check block: \\\hspace{3mm}``{\tt\textcolor{codegreen}{if (var != null) \{}buggy code\textcolor{codegreen}{\}}}'';\\
													2. Insert if-null-check block before buggy code: \\\hspace{3mm}
													   ``{\tt\textcolor{codegreen}{if (var == null) \{return false;\}} buggy code;}'' or \\\hspace{3mm}
													   ``{\tt\textcolor{codegreen}{if (var == null) \{return null;\}} buggy code;}'' or \\\hspace{3mm}
													   ``{\tt\textcolor{codegreen}{if (var == null) \{throw IllegalArgumentException.\}} buggy code;}''.}
													& \makecell[c]{C-4,26,\\C-14,19,\\{\em C-25},Cl-2,\\M-4,\\Moc-29,38.}\\\midrule
			{\tt DLS\_DEAD\_LOCAL\_STORE}        & \makecell[l]{Replace a variable with other one: \\e.g., ``{\tt var1 = \textcolor{red}{\sout{var2}}\textcolor{codegreen}{var3};}''.}  & \makecell[c]{C-11,24,\\L-6,57,59,\\M-33,59,T-7.}\\\midrule
			{\tt UC\_USELESS\_CONDITION}
			                                     & \makecell[l]{
													1. Mutate the operator of an expression in an {\tt if} statement:\\\hspace{2mm}
													   e.g., ``{\tt if (expA \textcolor{red}{\sout{>}}\textcolor{codegreen}{>=} expB) \{...\}}'';\\
													2. Remove a sub-predicate expression in an {\tt if} statement: \\\hspace{3mm}``{\tt if (\textcolor{red}{\sout{expA ||}} expB) \{...\}}'' or ``{\tt if (expA \textcolor{red}{\sout{|| expB}}) \{...\}}'';\\
													3. Remove the conditional expression: \\\hspace{3mm}
													   ``{\tt \textcolor{red}{\sout{expA ? expB :}} expC}'' or
													   ``{\tt \textcolor{red}{\sout{expA ?}} expB \textcolor{red}{\sout{: expC}}}''.}
												 & \makecell[c]{Cl-18,31,\\Cl-38,62,\\Cl-63,73,\\{\em L-15},M-46,\\M-82,85,\\T-19.}\\\midrule
			\protect{\tt UCF\_USELESS\_CONTROL\_FLOW}\protect\footnotemark{}
			                                     & \makecell[l]{
													1. Remove an {\tt if} statement but keep the code inside its block:\\``{\tt \textcolor{red}{\sout{if (exp) \{}} code \textcolor{red}{\sout{\}}}}'';\\
													2. Remove an {\tt if} statement with its block code:\\``{\tt \textcolor{red}{\sout{if (exp) \{ code \}}}}''.}
			                                     & \makecell[c]{{\em C-18}, {\em Cl-106},\\Cl-115,126,\\L-7,10,\\M-50.}\\\midrule
			{\tt UPM\_UNCALLED\_PRIVATE\_METHOD} & \makecell[l]{
													Remove a method declaration: \\``{\tt \textcolor{red}{\sout{Modifier ReutrnType methodName(Parameters) \{ code \}}}}''.}
												 & Cl-46,{\em M-77}.\\\midrule
			\protect{\tt BC\_UNCONFIRMED\_CAST}\protect\footnotemark{}
			                                     & \makecell[l]{Wrap buggy code with if-instanceof-check block: \\``{\tt\textcolor{codegreen}{if (var instanceof Type) \{}buggy code\textcolor{codegreen}{\}}}\\
			                                         {\tt\textcolor{codegreen}{else \{throw IllegalArgumentException;\}}}''.}
			                                     & M-89.\\
			\bottomrule
		\end{tabular}
		{\footnotesize $^\dagger$ Only correctly fixed (including partially correctly-fixed bugs highlighted with {\em italic}) bugs are listed in this table.
		}
	\end{threeparttable}
	}
\end{table*}


\begin{table*}[!t]
	\centering
	\caption{Dissection of bugs correctly fixed by \toolname.}
	\vspace{-1.5mm}
	\label{tab:fixedBugsDissection}
	{\scriptsize%
	\begin{threeparttable}
		\begin{tabular}{L{55mm}|L{50mm}|L{61mm}}
			\toprule
			{\bf Bug IDs} & {\bf Bug Patterns} & {\bf Repair Actions} \\
			\hline\hline
			C-1. & Conditional expression modification & Logic expression modification. \\
			\midrule
			C-4, 26. & Missing non-null check addition & Conditional (if) branch addition. \\
			\midrule
			C-14, 19, 25, Cl-2, M-4, Moc-29, 38. & Missing null check addition	 & Conditional (if) branch addition. \\
			\midrule
			C-11, 24, L-6, 57, 59, M-33, 59, T-7.  & Wrong variable reference & Variable replacement by another variable.\\
			\midrule
			Cl-38. & Logic expression expansion & Conditional expression expansion.\\
			\midrule
			Cl-18, 31, L-15. & Logic expression reduction & Conditional expression reduction.\\\midrule
			Cl-62, 63, 73, M-82, 85, T-19. &Logic expression modification	 & Conditional expression modification.\\
			\midrule
			C-18, Cl-106, M-46. & Unwraps-from if-else statement & Conditional (if or else) branch removal.\\
			\midrule
			Cl-115, 126, L-7, 10, M-50. & Conditional block removal & Conditional (if or else) branch removal. \\
			\midrule
			Cl-46, M-77. & Unclassified & Method definition removal.\\
			\midrule
			M-89. & Wraps-with if-else statement	 & Conditional (if-else) branches addition.\\
			\bottomrule
		\end{tabular}
	\end{threeparttable}
	}
\end{table*}

\subsection{Dissecting the Fix Ingredients}
\label{sec:dissection}

We now investigate how fix patterns of static analysis violations can be leveraged to address semantic bugs from the Defects4J benchmark. To that end, we dissect the ingredients that were successfully leveraged in the generated correct patches. Table~\ref{tab:fixedBugsFP} provides the summary of this dissection.

First, we note that all correctly fixed bugs were addressed with patches generated from patterns mined in the study of Liu et al.~\cite{liu2019mining} (i.e., based on FindBugs violations). Fix patterns from the study by Rolim et al.~\cite{rolim2018learning} (which are based on PMD violations) are indeed associated to exceedingly simple violations, which are unlikely to be revealed as semantic bugs (i.e., detected via developer test cases). An example of such simple pattern is their EP7 fix pattern: ``replace {\tt List$<$String$>$ a = new ArrayList()} with {\tt List$<$String$>$ a = new ArrayList$<>$()}''.
In any case, 6 among the 9 fix patterns released by Rolim et al. are related to performance, code practice or code style. Our manual investigation of Defects4J bugs reveals that none of the bugs are associated to these types of issues.

Second, we note that the fix patterns of only seven (out of 10) violation types have been successfully used to generate correct patches for Defects4J bugs (c.f. Table~\ref{tab:fixedBugsFP}).
Among the 40 (fully or partially) correctly fixed bugs, 36 (90\%) are fixed with fix patterns of 4 violation types: {\tt NP\_ NULL\_ON\_SOME\_PATH}, {\tt DLS\_DEAD\_LOCAL\_STORE}, {\tt UC\_ USELESS\_CONDITION}, and {\tt UCF\_USELESS\_CONTROL\_ FLOW}. The latter two violation types are related to the issues of conditional code entities (e.g., If statements and conditional expressions), which are relevant to 18 (45\%) of the fixed bugs. Comparing against the ACS~\cite{xiong2017precise} state-of-the-art APR tool which focuses on repairing condition-related faulty code entities, we find that \toolname correctly fixes 15 relevant bugs that are not fixed by ACS.

Finally, we investigate the diversity of the bugs that are correctly addressed by \toolname. To that end, we resort to the dissection study of Defects4J bugs by Sobreira et al.~\cite{sobreira2018dissection}. Table~\ref{tab:fixedBugsDissection} summarizes the bug patterns and the associated repair actions for the bugs that are correctly fixed by \toolname. We note that \toolname can currently address 11 bug patterns out of the 60 bug patterns enumerated in the dissection study.

\find{{\bf RQ2$\blacktriangleright$}\toolname exploits a variety of fix ingredients
from static violations fix patterns to address a diverse set of bug patterns in the Defects4J dataset.}

\vspace{-1mm}
\subsection{Comparing against the State-of-the-Art}
\label{sec:comparison}
To reliably compare against the state-of-the-art in Automated Program Repair (APR), we must ensure that the Fault Localization (FL) step is properly tuned as FL could bias the bug fixing performance of APR tools~\cite{liu2019you}. We identify three major configurations in the literature:

\begin{enumerate}[leftmargin=*]
	\item {\bf {\em Normal\_FL}-based APR}: in this case, APR systems directly use off-the-shelf fault localization techniques to localize the faulty code positions. In this case, which is realistic, the suspicious list of fault locations may be inaccurate leading to a poor repair performance. APR tools that are run in ASTOR~\cite{martinez2016astor} work under this configuration.
	\addtocounter{footnote}{-2}
	\stepcounter{footnote}\footnotetext{The current code contains a useless control flow statement, where the control flow continues onto the same place regardless of whether or not the branch is taken.}
	\stepcounter{footnote}\footnotetext{The cast expression is unchecked or unconfirmed, and not all instances of the type casted from can be cast to the type it is being cast to. It needs to check that the program logic ensures that this cast will not fail.}	
	\vspace{-4mm}
	\item {\bf {\em Restricted\_FL}-based APR}: in this case, APR systems make the assumption that some information of the code location is available. For example, in HDRepair~\cite{le2016history}, fault localization is restricted to the faulty methods, which are assumed to be known. Such a restriction actually substantially increases the accuracy of the target list of fault locations for which a patch must be generated. In Section~\ref{sec:frb}, we have made a similar strong assumption that fault locations are known as our objective was to assess the patch generation performance of \toolname.
	\item {\bf {\em Supplemented\_FL}-based APR}: in this case, APR systems
	leverage existing fault localization tools but improve the localization
	approach with some heuristics to ensure that the patch generation targets
	an accurate list of code locations. For example,
	the SimFix~\cite{jiang2018shaping} recent state-of-the-art system employs a test
	purification~\cite{xuan2014test} technique to improve the accuracy of the fault localization.
\end{enumerate}

We thus compare the bug fixing performance of \toolname with the state-of-the-art APR tools after classifying them into these three groups.

\subsubsection{Comparison against a Restricted\_FL-based APR System}

We first compare \toolname against the HDRepair~\cite{le2016history} state-of-the-art APR system, which implements a {\em restricted} fault localization configuration.
We select faulty locations using the same assumption as HDRepair, i.e., focusing on attempting to repair suspicious code statements that are reported by our fault localization tool but filtering only those that are within the known faulty methods.
This assumption leaves out many noisy statements, reducing the probability of generating overfitting patches for bugs and further increasing the chance to generate a correct patch before a plausible one or any execution timeout.


\begin{table}[!t]
	\centering
	\vspace{1mm}
	\caption{Comparison of \toolname with HDRepair~\cite{le2016history}.}
	\vspace{-1mm}
	\label{tab:compHDR}
	\begin{threeparttable}
		\begin{tabular}{L{12mm}|C{20mm}|C{18mm}|C{18mm}}
			\toprule
			\multirow{2}{*}{\bf Project} & \multirow{2}{*}{\bf HDRepair} & \multicolumn{2}{c}{\bf \toolname} \\\cline{3-4}
			 & & Fully fixed & Partially fixed\\
			\hline
			Chart       & 0/2 & 7/9  & 1/3 \\
			Closure     & 0/7 & 9/15 & 1/2 \\
			Lang        & 2/6 & 5/12 & 1/3 \\
			Math        & 4/7 & 6/14 & 1/3 \\
			Mockito     & 0/0 & 2/2  & 0/0 \\
			Time        & 0/1 & 2/3  & 0/0 \\
			\hline
			Total       & 6/23 & 31/55$^{\ast}$ & 4/11$^{\ast}$\\
			\hline
			P (\%)       & 26.1 & 56.4 & 36.4 \\
			\bottomrule
		\end{tabular}
		{\footnotesize The results for HDRepair are provided by its author. \\$^{\ast}$The number of bugs fixed by \toolname shown in this table is a little different from the data in Table~\ref{tab:fixedBugs}. For fixing each bug, the input of \toolname is a ranked list of suspicious statements in the faulty methods, which is different from the input of \toolname in the experiment of Section~\ref{sec:frb}.}
	\end{threeparttable}
\end{table}
\begin{table*}[!t]
	\centering
	\setlength\tabcolsep{2pt}
	\caption{Number of bugs reported as having been fixed by different APR systems.}
	\vspace{-1mm}
	\label{tab:comparison}
	\resizebox{0.8\linewidth}{!}
    {
    	\begin{threeparttable}
		\begin{tabular}{l|l|r|C{11mm}|C{11mm}|C{11mm}|C{11mm}|C{11mm}|C{11mm}|C{11mm}|C{11mm}}
			\toprule
			{\bf Fault Localization} & \multicolumn{2}{l|}{{\bf APR Tool}} & {\bf Chart} & {\bf Closure} & {\bf Lang} & {\bf Math} & {\bf Mockito} & {\bf Time} & {\bf Total} & {\bf P$^{\ast}$(\%)} \\
			\hline
			\multirow{8}{*}{{\em Normal\_FL}-based APR}
			& \multirow{2}{*}{\bf \toolname} & Fully Fixed & \cellcolor{grey}{\bf  5/12} & \cellcolor{grey}{\bf8/12} & \cellcolor{grey}{5/11} & \cellcolor{grey}{6/13} & \cellcolor{grey}{\bf 2/2} &\cellcolor{grey}{1/3} & \cellcolor{grey}{27/53}$^{\ast}$ & \cellcolor{grey}{50.9}\\
			\cline{3-11}
			& & Partially Fixed & \cellcolor{grey}{1/2} & \cellcolor{grey}{1/1} & \cellcolor{grey}{0/2} & \cellcolor{grey}{1/3} & \cellcolor{grey}{0/0} &\cellcolor{grey}{0/0} & \cellcolor{grey}{3/8}$^{\ast}$ & \cellcolor{grey}{37.5}\\
			\cline{2-11}
			& \multicolumn{2}{l|}{\bf jGenProg~\cite{martinez2016astor}} & 0/7 & 0/0 & 0/0 & 5/18 & 0/0 & 0/2 & 5/27 & 18.5\\
			& \multicolumn{2}{l|}{\bf jKali~\cite{martinez2016astor}} & 0/6 & 0/0 & 0/0 & 1/14 & 0/0 & 0/2 & 1/22 & 4.5 \\
			& \multicolumn{2}{l|}{\bf jMutRepair~\cite{martinez2016astor}} & 1/4 & 0/0 & 0/1 & 2/11 & 0/0  & 0/1 & 3/17 & 17.6 \\
			& \multicolumn{2}{l|}{\bf Nopol~\cite{xuan2017nopol}} & 1/6 & 0/0 & 3/7 & 1/21 & 0/0 & 0/1 & 5/35 & 14.3 \\
			& \multicolumn{2}{l|}{\bf FixMiner~\cite{koyuncu2018fixminer}} & 5/8 & 5/5 & 2/3 & 12/14 & 0/0 & 1/1 & 25/31 & 80.65 \\
			& \multicolumn{2}{l|}{\bf LSRepair~\cite{kui2018live}} & 3/8 & 0/0 & 8/{\bf14} & 7/14 & {1}/1 & 0/0 & 19/37 & 51.4 \\
			\hline
			\hline
			\multirow{7}{*}{{\em Supplemented\_FL}-based APR} &
			\multicolumn{2}{l|}{\bf ACS~\cite{xiong2017precise}} & 2/2 & 0/0 & 3/4 & 12/16 & 0/0 & 1/1 & 18/23 & 78.3 \\
			& \multicolumn{2}{l|}{\bf ELIXIR~\cite{saha2017elixir}} & 4/7 & 0/0 & 8/12 & 12/19 & 0/0 & {\bf2}/3 & 26/41 & 63.4 \\
			& \multicolumn{2}{l|}{\bf JAID~\cite{chen2017contract}} & 2/4 & 5/{11} & 1/8 & 1/8 & 0/0 & 0/0 & 9/31 & 29.0 \\
			& \multicolumn{2}{l|}{\bf ssFix~\cite{xin2017leveraging}} & 3/7 & 2/{11} & 5/12 & 10/{\bf26} & 0/0 & 0/{\bf4} & 20/{\bf60} & 33.3 \\
			& \multicolumn{2}{l|}{\bf CapGen~\cite{wen2018context}} & 4/4 & 0/0 & 5/5 & 12/16 & 0/0 & 0/0 & 21/25 & {\bf84.0} \\
			& \multicolumn{2}{l|}{\bf SketchFix~\cite{hua2018towards}} & {\bf 6}/8 & 3/5 & 3/4 & 7/8 & 0/0 & 0/1 & 19/26 & 73.1 \\
			& \multicolumn{2}{l|}{\bf SimFix~\cite{jiang2018shaping}} & 4/8 & {6}/8 & {\bf9}/13 & {\bf14}/{\bf26} & 0/0 & 1/1 & {\bf34}/56 & 60.7 \\
			\bottomrule
			\end{tabular}
			{``{\bf P}'' is the probability of generated plausible patches to be correct. \\$^{\ast}$The number of bugs fixed by \toolname shown in this table is a little different from the data in Table~\ref{tab:fixedBugs} and Table~\ref{tab:compHDR}. In this experiment, for fixing each bug, the input of \toolname is a ranked full list of suspicious statements in the faulty program, which is different from the input of \toolname in the experiments of Table~\ref{tab:fixedBugs} and Table~\ref{tab:compHDR}.
			}
		\end{threeparttable}
	}
\end{table*}

\begin{figure}[!h]
	\centering
    \includegraphics[width=\linewidth]{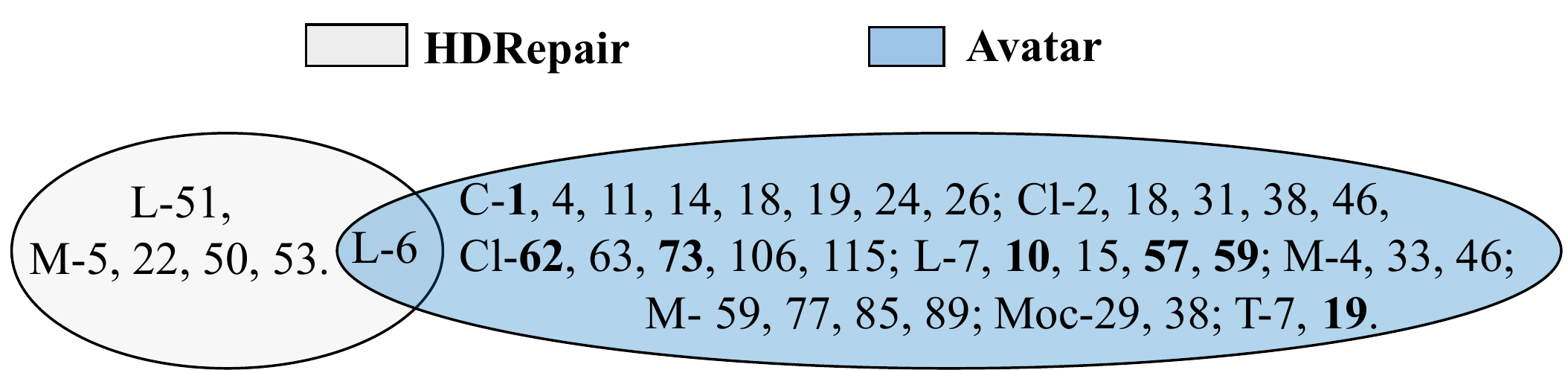}
    \caption{Bugs correctly fixed by HDRepair and \toolname, respectively.}
    \label{fig:avatar_hdrepair}
\end{figure}

Table~\ref{tab:compHDR} presents the comparing results. Comparing with HDRepair, \toolname correctly fixes many more bugs (31+4 vs 6) and yields a higher probability to generate correct patches among all plausible patches (cf. P(\%) in Table~\ref{tab:compHDR}). 34 (except {\em Lang-6}) out of the 35 bugs fixed by \toolname are not addressed by HDRepair. \toolname also correctly fixes 7 bugs (as highlighted with {\bf bold} in Figure~\ref{fig:avatar_hdrepair}) that are only plausibly (but incorrectly) fixed by HDRepair.
Finally, \toolname partially fixes 11 bugs that have multiple faulty code fragments, and 4 of the associated patches are correct. Figure~\ref{fig:avatar_hdrepair} illustrates the space of correctly fixed bugs by HDRepair and \toolname.

\find{{\bf RQ3$\blacktriangleright$}\toolname substantially outperforms the HDRepair approach on the Defects4J benchmark.}

\vspace{-4mm}
\subsubsection{Comparison against Normal\_FL-based APR Systems}

We compare the bug fixing performance of \toolname with the {\em Normal\_FL}-based state-of-the-art APR tools that are evaluated on the Defects4J benchmark.
These APR tools take as input a ranked list of suspicious statements that are reported by an off-the-shelf fault localization technique.
In this experiment, we consider a group of APR systems, namely jGenProg~\cite{martinez2017automatic}, jKali~\cite{martinez2017automatic}, jMutRepair~\cite{martinez2016astor}, Nopol~\cite{xuan2017nopol}, FixMiner~\cite{koyuncu2018fixminer} and LSRepair~\cite{kui2018live}, which leverage a similar configuration as \toolname for fault localization: GZoltar/Ochiai.

Table~\ref{tab:comparison} reports the comparison results in terms of
the number of {\em plausibly-fixed} bugs and the number of {\em correctly-fixed} bugs.
Data about the fixed bugs are directly excerpted from the results reported
in the relevant research papers. We note that \toolname outperforms all of
the {\em Normal\_FL}-based APR systems, both in terms of the number of plausibly
fixed bugs and the number of correctly fixed bugs. It also yields
a higher probability to generate correct patches among its plausible patches
than those tools (except FixMiner). Finally, 18 (15 + 3, as shown in
the second row of Table~\ref{tab:unfixedBugs}) among the 30 (27 + 3) bugs correctly
fixed by \toolname have not been correctly fixed by those {\em Normal\_FL}-based APR tools.

\find{{\bf RQ3$\blacktriangleright$}In terms of quantity and quality of generated plausible patches, \toolname addresses more bugs than its immediate competitors. Nevertheless, we note that \toolname is actually complementary to the other state-of-the-art APR systems, fixing bugs that others do not fix.}

\subsubsection{Comparison against Supplemented\_FL-based APR Systems}
We also compare \toolname against APR systems which use supplementary information to improve fault localization accuracy. We include in this category other APR systems whose authors do not explicitly describe the actual fault localization configuration, but which still manage to fix bugs that we could not localize with GZoltar/Ochiai.
We include in this group the following state-of-the-art works targeted at Java programs: ACS~\cite{xiong2017precise}, ELIXIR~\cite{saha2017elixir}, JAID~\cite{chen2017contract}, ssFix~\cite{xin2017leveraging}, CapGen~\cite{wen2018context}, SketchFix~\cite{hua2018towards} and SimFix~\cite{jiang2018shaping}.

\begin{table*}[!h]
	\centering
	\caption{Bugs fixed by \toolname but not correctly fixed by other APR tools.}
	\vspace{-1mm}
	\label{tab:unfixedBugs}
	{%
	\begin{threeparttable}
		\begin{tabular}{l|c|c}
			\toprule
			\multirow{2}{*}{\bf APR tool group} & \multicolumn{2}{c}{\bf Bug IDs}\\\cline{2-3}
			& Fully-fixed & Partially-fixed \\
			\hline
			{\em Normal\_FL}-based APR tools & C-14,19,Cl-2,18,31,46,L-6,7,10,M-4,46,59,Moc-29,38,T-7. & C-18, Cl-106, M-77.\\\midrule
			{\em Supplemented\_FL}-based APR tools & C-4,Cl-2,31,38,46,L-6,7,10,M-46,Moc-29,38. & C-18, Cl-106, M-77.\\\midrule
			All APR tools                    & 	Cl-2,31,46,L-7,10,M-46,Moc-29,38. &  C-18, Cl-106, M-77.\\
			\bottomrule
		\end{tabular}
	\end{threeparttable}
	}
\end{table*}

The compared performance results are also illustrated in Table~\ref{tab:comparison}. Based on the number of
correctly fixed bugs, \toolname is only inferior to SimFix but outperforms other {\em
Supplemented\_FL}-based APR systems.
\toolname further correctly fixes 14 (11 + 3, as shown in the third row of Table~\ref{tab:unfixedBugs}) out
of 31 bugs that have never been addressed by any {\em Supplemented\_FL}-based state-of-the-art APR system.

To sum up, \toolname correctly fixes 11 (as shown in the fourth row of Table~\ref{tab:unfixedBugs}) out of
31 bugs that have never been addressed by any state-of-the-art APR system. We also note that \toolname
outperforms state-of-the-art APR tools on fixing bugs in project
{\em Chart}, {\em Closure} and {\em Mockito}.

\find{{\bf RQ3$\blacktriangleright$}\toolname underperforms against some of the most recent APR systems. Nevertheless, \toolname is still complementary to them as it is capable of addressing some Defects4J bugs that the state-of-the-art cannot fix.}

\section{Threats to Validity}
\label{sec:dis}

A threat to external validity is related to use of Defects4J bugs as a representative set of semantic bugs. This threat is mitigated as it is currently a widely used dataset in the APR literature related to Java.
A threat to internal validity is due to the use of Java programs as subjects. Eventually, we only considered fix patterns for FindBugs and PMD violations. Other static tools, especially for C programs, such as Splint, cppcheck,
and Clang Static Analyzer are not investigated. 
A threat to construct validity may involve the assumption of perfect localization to assess \toolname. This threat is minimized by the different other experiments that are comparable with evaluations in the literature.

\section{Related Work}
\label{sec:relatedWork}
The software development practice is increasingly accepting generated patches~\cite{koyuncu2017impact}. Recently, various directions in the literature have been explored to contribute to the advancement of automated program repair.
One commonly studied direction is the pattern based (also called example-based) APR. Kim et al.~\cite{kim2013automatic} initiated with PAR a milestone of APR  based on fix templates that were manually extracted from 60,000 human-written patches. Later studies~\cite{le2016history} have shown that the six templates used by PAR could fix only a few bugs in Defects4J.
Long and Rinard also proposed a patch generation system, Prophet~\cite{long2016automatic}, that learns code correctness models from a set of successful human patches. They further proposed a new system, Genesis~\cite{long2017automatic}, which can automatically infer patch generation transforms from developer submitted patches for program repair.

Motivated by PAR~\cite{kim2013automatic}, more effective automated program repair systems have been explored. HDRepair~\cite{le2016history} was proposed to repair bugs by mining closed frequent bug fix patterns from graph-based representations of real bug fixes. Nevertheless, its fix patterns, except the fix templates from PAR, still limits the code change actions at abstract syntax tree (AST) node level, but are not specific for some types of bugs.
ELIXIR~\cite{saha2017elixir} aggressively uses method call related templates from PAR with local variables, fields, or constants,
to construct more expressive repair-expressions that go into synthesizing patches.

Tan et al.~\cite{tan2016anti} integrated anti-patterns into two existing search-based automated program repair tools (namely, GenProg~\cite{le2012genprog} and SPR~\cite{long2015staged}) to help alleviate the problem of incorrect or incomplete fixes resulting from program repair. In their study, the anti-patterns are defined by themselves and limited to the control flow graph.
Additionally, their anti-patterns are not meant to solve the problem of deriving better patches automatically, provide more precise repair hints to developers.

More recently, CapGen~\cite{wen2018context}, SimFix~\cite{jiang2018shaping}, FixMiner~\cite{koyuncu2018fixminer} are further proposed to fix bugs automatically based on the frequently occurred code change operations (e.g., Insert IfStatement (c.f., Table 3 in~\cite{jiang2018shaping}) that are extracted from the patches in developer change histories. 

So far however, pattern-based APR approaches focus on leveraging patches that developer applied to semantic bugs. To the best of our knowledge, our approach is first to investigate the case of leveraging patches that fix static analysis violations: they are many more, better identifiable, and more consistent.


\section{Conclusion}
\label{sec:conc}
The correctness of patches generated is now identified as a barrier in the adoption of automated program repair systems. 
Towards guaranteeing correctness, researchers have been investigating example-based approaches where fix patterns from human patches are leveraged in patch generation. Nevertheless, such ingredients are often hard to collect reliably. 
In this work, we propose to rely on developer patches that address static analysis bugs. 
Such patches are concise and precise, and their efficacy (in removing the bugs) are systematically assessed (by the static detectors). 
We build \toolname, an APR system that utilizes fix ingredients from static analysis violations patches. 
We empirically show that \toolname is indeed effective in repairing programs that have semantic bugs. \toolname outperforms several state-of-the-art approaches and complements others by fixing some of the Defects4J bugs which were not yet fixed by any APR system in the literature.

As future work, we plan to assess avatar on bigger bug datasets~\cite{saha2018bugs}, and use this concept of static analysis-based patterns for improving method refactoring research~\cite{pradel2018deepbugs, liu2019learning}.

\section*{Acknowledgements}
{
This work is supported by the Fonds National de la Recherche (FNR), Luxembourg, through RECOMMEND 15/IS/10449467 and FIXPATTERN C15/IS/9964569.
}

\balance
\bibliographystyle{IEEEtran}
\bibliography{bib/references}

\end{document}